\documentstyle[11pt]{article}

 \newcommand{\beq}{\begin{equation}}
\newcommand{\eeq}{\end{equation}} \newcommand{\beqn}{\begin{eqnarray}}
\newcommand{\eeqn}{\end{eqnarray}} 
\newcommand{\balpha}{{\mbox{\boldmath $\alpha$}}}

\newcommand{\bbeta}{{\mbox{\boldmath $\beta$}}}

\begin{document}
\thispagestyle{empty}
\baselineskip=18pt
\rightline{UOSTP-99-006}
\rightline{SNUTP-99-034}
\rightline{KIAS-P99053}
\rightline{{\tt hep-th/9907090}}
\vskip 2cm
\centerline{\LARGE\bf Quantum 1/4 BPS Dyons}

\vskip 0.2cm

\vskip 1.2cm
\centerline{\large\it
Dongsu Bak $^a$\footnote{Electronic Mail: dsbak@mach.uos.ac.kr},
Kimyeong Lee $^b$\footnote{Electronic Mail: kimyeong@phya.snu.ac.kr},
and Piljin Yi $^c$\footnote{Electronic Mail: piljin@kias.re.kr}}
\vskip 10mm
\centerline{ \it $^a$ Physics Department,
University of Seoul, Seoul 130-743, Korea}
\vskip 3mm
\centerline{ \it $^b$ Physics Department and Center for Theoretical
Physics}
\centerline{ \it Seoul National University, Seoul 151-742, Korea}
\vskip 3mm
\centerline{ \it $^c$ School of Physics, Korea Institute for
Advanced Study}
\centerline{\it
207-43, Cheongryangri-Dong, Dongdaemun-Gu, Seoul 130-012, Korea}

\vskip 1.2cm
\begin{quote}
{\baselineskip 16pt
Classical properties of 1/4 BPS dyons were previously well understood both 
in the field theory context and in the string theory context. Its quantum
properties, however, have been more difficult to probe, although the
elementary information of the supermultiplet structures is known
from a perturbative construction. Recently, a low energy effective theory
of monopoles was constructed and argued to contain these dyons as quantum
bound states. In this paper, we find these dyonic bound states explicitly
in the $N=4$ supersymmetric low energy effective theory. After identifying
the correct angular momentum operators, we motivate an anti-self-dual
ansatz for all BPS bound states. The wavefunctions are found explicitly,
whose spin contents and degeneracies match exactly the expected results.

}

\end{quote}


\newpage
\section{Introduction}

In the N=4 supersymmetric Yang-Mills theories, there can be 1/2 BPS
and 1/4 BPS configurations. The precise nature of these
states depends on the asymptotic values of six Higgs fields
in the theory. When the Higgs expectation values have only one
independent component, only 1/2 BPS configurations can appear.
Classically, 1/2 BPS configurations are made of monopoles or dyons,
and its electric field is proportional to its magnetic field.
When the Higgs expectation values have two or more independent
components, 1/4 BPS configurations can also appear, which are all
dyons whose electric charge is not proportional to its magnetic charge
\cite{bergman,yi,hash}.

Any 1/2 BPS configuration of a given collection of monopoles is
specified by their moduli parameters, and the low energy dynamics
of these 1/2 BPS monopoles is determined by the metric on the manifold
spanned by these moduli parameters \cite{manton}.
On the other hand, as shown in Ref.~\cite{yi,hash}, solutions
to 1/4 BPS equations can be obtained in two steps: First, one
solves for purely magnetic soliton, which may be regarded as 1/2 BPS
in a technical sense. Then, one solves for certain linear combination of
gauge zero modes in this purely magnetic background. Curiously enough,
one can build the electric part of the 1/4 BPS dyons from such gauge zero
mode. Because existence of gauge zero modes are guaranteed for any 1/2
BPS monopole \cite{ejw},
any 1/4 BPS configuration is again specified by the moduli
parameters of the corresponding 1/2 BPS monopole configuration.
Given a fixed set of moduli parameters, the  solution of the second
BPS equation determines relative part of the electric charge of 
monopoles uniquely.

One may regard 1/4 BPS configurations as deformed 1/2 BPS configurations 
when the additional and independent Higgs expectation is turned on.
When this second Higgs expectation  is quite small
compared to the first, the
deviation of the 1/4 BPS configurations from the 1/2 BPS configuration is
small. In such cases, the authors (with C. Lee) have shown in a recent
paper \cite{blly} that one can describe the low energy dynamics of both
1/2 BPS and 1/4 BPS configurations with an effective nonrelativistic 
Lagrangian.

The kinetic part of the Lagrangian is given by the moduli space metric of
the 1/2 BPS configurations. The potential is also present, and is given by
the square of the norm of a triholomorphic Killing vector field related to
an unbroken $U(1)$ gauge symmetry. The size of this attractive potential
is proportional to the square of the additional Higgs expectation value.
This effective Lagrangian can be interpreted as low energy dynamics of
1/2 BPS monopoles with attractive potential, in other words, and the 
1/4 BPS configurations should be realized as BPS bound states of monopoles 
with additional  electric quantum numbers.

In Ref.~\cite{blly},
the full N=4 supersymmetric low energy effective Lagrangian
is written. This is a sigma model with potential that has extended complex
supersymmetry with a central term, and, as usual, the wavefunctions can be interpreted as
differential forms on the moduli space. The BPS equation was found and
translated to the language of differential forms.

The simplest nontrivial 1/4 BPS configurations appear as composite
of two distinct fundamental magnetic monopoles in $SU(3)$ gauge
theory. Furthermore, the 8-dimensional moduli space of these two monopoles
is known exactly. From Ref.~\cite{yi},
several facts are known about this case.
First of all, classical 1/4 BPS configurations are made of two 1/2 BPS dyons
at rest, whose mutual distance is determined by their relative
electric charge. Also, the supermultiplet structures of all such dyons
have been found by the perturbative method around the 1/2 BPS state of
the zero relative electric charge, where the nonzero relative charge
states are constructed by exciting certain massive excitations on
1/2 BPS configurations.

In this paper, we reconstruct these 1/4 BPS dyons as quantum bound state
of two distinct $SU(3)$ monopoles in the low energy dynamics described
above. We construct all such $SU(3)$ dyons. We also recover
the phenomenon of instability found in Ref.~\cite{bergman,yi}:
The bound state wavefunction
loses its normalizability exactly at the point where the instability
should set in. Furthermore, we explicitly show that each (stable) dyon
comes in the same supermultiplet as found in Ref.~\cite{yi}

The plan of the paper is as follows. In Sec. 2, we briefly discuss the
moduli space of a pair of distinct monopoles in $SU(3)$ theory. In Sec. 3,
we review briefly the supersymmetric Hamiltonian and BPS conditions on 
wavefunctions as shown in Ref.~\cite{blly}.
In Sec. 4, we discuss the angular momentum
and an ansatz for the BPS wave functions. In Sec. 5, we solve the BPS
equations. In Sec. 6, we conclude with some comments.

\section{A Pair of Distinct Monopoles in the SU(3) Gauge Theory}

Consider $N=4$ $SU(3)$ gauge theory spontaneously broken to $U(1)^2$.
When the six Higgs expectations are all collinear, the theory contains
two distinct types of fundamental monopoles, which  we will label
by $\balpha$ and $\bbeta$. The low energy interaction between $\balpha$
and $\bbeta$ monopoles can be described by the moduli space dynamics.
There are four collective coordinates for each monopole, three for its
position and one for the $U(1)$ phase. We call their
positions and phases to be ${\bf x}_i,\chi_i, i=1,2$, for $\balpha$ and
$\bbeta$ monopoles, respectively.  Let us parameterize the masses of these
monopoles as $\mu_1$ and $\mu_2$. We are suppressing the gauge coupling 
constant in all subsequent formulae.

The exact nonrelativistic effective Lagrangian has been found to be a sum
of the Lagrangians for the center of mass and the relative motion
\cite{atiyah}. As there is no
external force, the center of mass Lagrangian is a free one:
\beq
{\cal L}_{\rm cm} =\frac{(\mu_1+\mu_2)}{2} \dot{{\bf X}}^2 +
\frac{1}{2(\mu_1+\mu_2)} \dot{\chi}_T^2 ,
\label{cmL}
\eeq
where the center of mass position is ${\bf X} = ( \mu_1 {\bf x}_1 +
\mu_2 {\bf x}_2)/(\mu_1+\mu_2)$ and the center of mass phase is $\chi_T
=\chi_1+\chi_2 $.
The relative motion between them is more complicated and described
by the Taub-NUT metric \cite{connell,lwy}, and has the Lagrangian,
\beq
{\cal L}_{\rm rel}
= \frac{\mu}{2} \left( ( 1+ \frac{1}{\mu r}) \dot{\bf r}^2
+ \frac{1}{\mu^2(1+ \frac{1}{\mu r})}
(\dot{\chi} + {\bf w}({\bf r})\cdot \dot{\bf r})^2 \right),
\label{relL}
\eeq
where the relative position is ${\bf r} = {\bf x}_2-{\bf x}_1$, the
relative phase is $\chi =2(\mu_1\chi_2-\mu_2\chi_1)/(\mu_1+\mu_2)$,
and ${\bf w}({\bf r}) $ is the Dirac potential such that $\nabla
\times {\bf w}({\bf r}) = -{\bf r}/r^3$. The range of $\chi$ is
$[0,4\pi]$. From now on, we will suppress the scale $\mu$ by setting
$\mu=1$. The resulting monopole moduli space metric is then,
\beq
{g}_{\rm rel}
=   \left( 1+ \frac{1}{r}\right) d{\bf r}^2
+ \left(\frac{1}{1+ {1}/{ r}}\right)
(d{\chi} + {\bf w}({\bf r})\cdot d{\bf r})^2 ,
\eeq
up to an overall scale.
This Taub-NUT space, ${\cal M}_0$, has
the topology of $R^4=R^+\times S^3$.
The eight-dimensional, total moduli space is then given by
\beq
{\cal M} = R^3 \times \frac{ R^1 \times {\cal M}_0}{Z},
\eeq
where $Z$ is the identification map
\beq
(\chi_T, \chi) = (\chi_T + 2\pi, \chi + \frac{4\pi \mu_2}{\mu_1+\mu_2}).
\eeq

For later convenience, we will make another choice of coordinates
involving Euler angles on $S^3$,
\begin{equation}
{g}_{\rm rel}=\left(1+\frac{1}{r}\right)\,[dr^2+r^2\sigma_1^2+r^2\sigma_2^2]+
\frac{1}{1+1/r}\,\sigma_3^2,
\end{equation}
where the $\sigma_a$'s are 1-form frames on $S^3$, and satisfy the canonical
relationship,
\begin{equation}
d\sigma_a=\frac{1}{2}\epsilon_{abc}\,\sigma_b\wedge\sigma_c.
\end{equation}
More explicitly, we may write these 1-forms in terms of $SU(2)$ Euler angles
as follows
\begin{eqnarray}
&&\sigma_1= -
\sin\chi d\theta +\cos\chi\sin\theta d\phi,\nonumber\\
&&\sigma_2=
\cos\chi d\theta +\sin\chi\sin\theta d\phi,\nonumber\\
&&\sigma_3= d\chi+\cos\theta d\phi.
\end{eqnarray}
The ranges of $\theta$, $\phi$, $\chi$ are respectively $\pi$, $2\pi$,
$4\pi$. Let us define an orthonormal basis $\omega^\mu$ by
\begin{eqnarray}
\omega^0&=&\sqrt{1+1/r}\,dr ,\nonumber \\
\omega^1&=&\sqrt{r^2+r}\,\sigma_1,\nonumber \\
\omega^2&=&\sqrt{r^2+r}\,\sigma_2,\nonumber \\
\omega^3&=&\sqrt{\frac{r}{1+r}}\,\sigma_3.
\end{eqnarray}
Because the Taub-NUT manifold is a hyperK\"ahler 4-manifold, its
curvature is anti-self-dual with an appropriate choice of orientation.

When the Higgs vacua is slightly misaligned, the two monopoles are attracted
to each other \cite{hollowood}. The effective low energy potential 
${\cal U}$ of this static force has been found in Ref.~\cite{blly}
for all multimonopole configurations in
all $N=4$ gauge theories. Specializing to the case of a pair of distinct
monopoles in $SU(3)$, the relative part  of this potential is given by a
squared
norm of the Killing vector field $\partial_\chi$, up to an overall factor,
\beq
{\cal U}_{\rm rel}=\frac{1}{2}a^2\left\langle \frac{\partial}{\partial\chi},
\frac{\partial}{\partial\chi}\right\rangle,
\eeq
where $a$ is a measure of Higgs misalignment. The interacting part of
the two monopole dynamics is dictated by an effective Lagrangian, whose
bosonic part is
\begin{equation}
{\cal L}_{\rm rel}=\frac{1}{2}\,(g_{\rm rel})_{\mu\nu}
\dot z^\mu\dot z^\mu-{\cal U}_{\rm rel}
=\frac{1}{2}\left( 1+ \frac{1}{r}\right) \dot{\bf r}^2+ \frac{1}{2}
\left(\frac{1}{1+ {1}/{ r}}\right)(\dot{\chi} + {\bf w}({\bf r})\cdot
\dot{\bf r})^2 -\frac{1}{2}\left(\frac{a^2}{1+1/r}\right) .
\end{equation}
Note that the potential ${\cal U}_{\rm rel}$ increases from zero at origin
to $a^2/2$ at infinity. This behavior allows new bound states of the
dynamics which would not have been possible for $a=0$. Among these
are certain dyonic states that preserve 1/4 of field theory supersymmetries.
The purpose of this note is to reconstruct these 1/4 BPS dyons as
BPS quantum bound states in the low energy dynamics of monopoles.

\section{Supersymmetry and BPS Bound}

We begin by recapitulating generic properties of
the N=4 supersymmetric quantum extension of the bosonic
effective action~\cite{blly,alvarez,blum}.
Its form is rather similar to the usual
supersymmetric sigma model action but supplemented by an attractive bosonic
potential together with its fermionic counter part. These potentials
are determined by a single Killing vector field $G$.
The supersymmetric Lagrangian written with real fermions
is
\beqn
{\cal L}&=&{1\over 2} \biggl( g_{\mu\nu} \dot{z}^\mu \dot{ z}^\nu +
ig_{\mu\nu} \bar\psi^\mu \gamma^0 D_t \psi^\nu + {1\over 6}
R_{\mu\nu\rho\sigma}\bar\psi^\mu \psi^\rho \bar\psi^\nu \psi^\sigma
\biggr.
\nonumber\\
&& \biggl. - g^{\mu\nu} G_\mu G_\nu - D_\mu G_\nu  \bar\psi^\mu
\gamma_5\psi^\nu  \biggr),
\label{action}
\eeqn
where  $\psi^\mu$  is a two-component anticommuting Majorana spinor
and $\gamma^0= \sigma_2$, $\gamma_5= \sigma_3$, and $\bar\psi=\psi^T
\gamma^0$. In case of relative dynamics of the two $SU(3)$ monopoles,
$G$ is equal to $a\partial_\chi$. As required for the N=4 supersymmetry,
the metric here is hyperK\"ahler, endowed with three complex
structures ${\cal I}^{(a)\mu}\,_\nu (a=1,2,3)$ that satisfy
\beqn
&&{\cal I}^{(a)}{\cal I}^{(b)}  =
- \delta^{ab} +\epsilon^{abc} {\cal I}^{(c)}, \\
&& D_\mu {\cal I}^{(a)\nu}\,_\rho  =0\, .
\label{complex}
\eeqn
For the sake of N=4 supersymmetry,
the Killing vector $G\equiv a\cdot K$ should  be
triholomorphic; namely its action
preserves the three complex structures via
\beq
{\cal L}_{G} {\cal I}^{(a)}=0,
\label{triholomorphic}
\eeq
where $\cal L$ denotes the Lie derivative.

Upon quantization,
the spinors $\psi^A = e_\mu^A \psi^\mu$ with vielbein $e_\mu^A$,
commute with all the bosonic dynamical variables, especially with
$p$'s that are canonical momenta of the coordinates $z$'s.
The remaining  canonical commutation relations are
\beqn
&&[z^\mu, p_\nu ] = i\delta^\mu_\nu, \nonumber\\
&&\{\psi^A_\alpha, \psi^B_\beta\} = \delta^{AB}\delta_{\alpha\beta}\,.
\label{commutators}
\eeqn
The Lagrangian (\ref{action}) is invariant under the N=4
supersymmetry transformations,
\beqn
&&
\delta_{(0)} z^\mu= \bar\epsilon \psi^\mu, \\
&& \delta_{(0)}\psi^\mu  = -i\dot{z}^\mu\gamma^0
\epsilon  - \Gamma^\mu_{\nu\lambda}
\bar\epsilon \psi^\nu \psi^\lambda -\gamma_5 G^\mu
\epsilon, \\
&&\delta_{(a)}z^\mu  =
{\cal I}^{(a)\mu}\,_\nu \,\bar\epsilon_{(a)} \psi^\nu ,\\
&& \delta_{(a)}({\cal I}^{(a)\mu}\,_\nu\psi^\nu)  =
-i\dot{z}^\mu\gamma^0\,
\epsilon_{(a)}  - \Gamma^\mu_{\nu\lambda}
{\cal I}^{(a)\nu}\,_\rho
 {\cal I}^{(a)\lambda}\,_\sigma
\,\bar\epsilon_{(a)} \psi^\rho \psi^\sigma -\gamma_5 G^\mu\,
\epsilon_{(a)}  \,,
\label{transformation}
\eeqn
where $\epsilon$ and $\epsilon_{(a)}$ are  spinor parameters.
In order to obtain supercharges, we
define  supercovariant momenta by
\begin{eqnarray}
&& \pi_\mu \equiv p_\mu -{i\over 2}\omega_{AB\mu}
\bar\psi^A \gamma^0 \psi^B,
\label{cov}
\end{eqnarray}
where $\omega^A\,_{B\mu}$ is the spin connection.
The corresponding N=4 SUSY generators in real form are then
\begin{eqnarray}
&&Q_\alpha = \psi^\mu_\alpha \pi_\mu
+i(\gamma^0\gamma_5 \psi^\mu)_\alpha G_\mu ,\\
&&Q^{(a)}_\alpha = {{\cal I}^{(a)}}^\mu\,_\nu \psi^\nu_\alpha \pi_\mu
+i(\gamma^0\gamma_5 {{\cal I}^{(a)}}^\mu\,_\nu\psi^\nu)_\alpha G_\mu,
\label{generator}
\end{eqnarray}
which satisfy the following SUSY algebra with a central extension:
\beqn
&&\{Q_\alpha,Q_\beta  \}  =\{Q^{(a)}_\alpha,Q^{(a)}_\beta\}=2
 \delta_{\alpha\beta} \; {\cal H} +
2 i(\gamma^0\gamma_5)_{\alpha\beta} \; {\cal Z}, \\
&& \{Q_\alpha,Q^{(a)}_\beta  \} =0 ,\\
&& \{Q^{(a)}_\alpha,Q^{(b)}_\beta  \} =0 \ \ \ (a\neq b).
\label{algebra}
\eeqn
The Hamiltonian $\cal H$ and the central charge $\cal Z$ read
\beqn
&&{\cal H}=
{1\over 2} \biggl( {1\over \sqrt{g}}\pi_\mu \sqrt{g }g^{\mu\nu}\pi_\nu
+ G_\mu G^\mu -{1\over 4}R_{\mu\nu\rho\sigma}\bar\psi^\mu
\gamma^0 \psi^\nu \bar\psi^\rho
\gamma^0 \psi^\sigma + D_\mu G_\nu  \bar\psi^\mu \gamma_5\psi^\nu \biggr),\\
&& {\cal Z}= G^\mu \pi_\mu -{i\over 2}  (D_\mu G_\nu) \bar\psi^\mu
\gamma^0\psi^\nu.
\label{hamiltonian}
\eeqn
It is  easily checked that the central charge $\cal Z$ indeed commutes with
all SUSY generators.

For spectrum analysis, SUSY generators in complex form are more useful.
Introducing $\varphi\equiv {1\over \sqrt{2}} (\psi_1^\mu -i\psi_2^\mu)$,
and defining $ Q\equiv {1\over \sqrt{2}} (Q_1-iQ_2)$,
one finds
\beqn
&&Q = \varphi^\mu \pi_\mu
+ i \varphi^{*\mu} G_\mu,\\
&& Q^{\dagger}=\varphi^{*\mu} \pi_\mu
- i \varphi^{\mu} G_\mu,
\label{comgenerator1}
\eeqn
which generates the following simple algebra:
\begin{eqnarray}
&&\{Q,Q^{\dagger}\}=\{Q^{(a)},{Q^{(a)}}^{\dagger}\}  =2 {\cal H},\\
&&\{Q,Q\}  = \{Q^{(a)},Q^{(a)}\}=-\{Q^{\dagger},Q^{\dagger}\} =
-\{{Q^{(a)}}^{\dagger},{Q^{(a)}}^{\dagger}\}=
2i{\cal Z},\\
&& \{Q,Q^{(a)}\}=\{Q^{\dagger}, Q^{(a)}\}=0, \\
&& \{Q^{(a)},Q^{(b)}\}=\{{Q^{(a)}}^{\dagger}, Q^{(b)}\}=0, \ \ \ (a\neq b)
\label{algebra2}
\end{eqnarray}

It is easy to read off  the  BPS condition for quantum states that preserves
half of supersymmetries. Depending on the sign of central charge, we find
\begin{eqnarray}
&&(Q\mp iQ^\dagger)|\Phi\rangle  =0,
\label{comgenerator2}
\end{eqnarray}
so that the given state may  saturate the condition ${\cal H}=\pm {\cal Z}$.
We can express this BPS condition in more geometrical fashion by transcribing
the wavefunction to differential forms on the moduli
space~\cite{witten1}. Note that
\beqn
&& [i\pi_\mu,\varphi^\nu]= -\Gamma^\nu_{\mu\rho} \varphi^{\rho},\\
&& [i\pi_\mu, \varphi^*_\nu]=
\Gamma^\rho_{\mu\nu} \varphi^*_{\rho},\\
&& \{ \varphi^\mu,\varphi^{*}_\nu\}=\delta_\nu^\mu.
\label{comgenerator3}
\end{eqnarray}
Furthermore, the wavefunction has the following general form,
\beqn
&& |\Phi \rangle =
\sum_p \frac{1}{p!}\,\Omega_{\mu_1 \cdots \mu_p}(z^\mu)
\varphi^{\mu_1}\cdots \varphi^{\mu_p}|0\rangle \\
&& \varphi^{*\mu}|0\rangle =0.
\label{comgenerator}
\eeqn
with a inner product defined by
\beqn
\langle \Phi  |\Phi' \rangle =
\int dx\sqrt{g}\sum_p (\Omega^{\mu_1 \cdots \mu_p})^*
\Omega'_{\mu_1 \cdots \mu_p}.
\eeqn
The coefficients $\Omega_{\mu_1 \cdots \mu_p}$ are completely antisymmetric
and may be regarded as those of a $p$-form. In this language where we
interpret $\varphi^\mu$ and $\varphi^*_\mu$ as
a natural cobasis $dz^\mu$ and a natural basis $\partial\over \partial
z^\mu$,  one finds that the following replacement can be made:
\beqn
&& i\varphi^\mu \pi_\mu \rightarrow d \,,
\ \ i\varphi^{*\mu} \pi_\mu \rightarrow -\delta,\\
&& \varphi^{*\mu} G_\mu \rightarrow i_G\,, \ \
 i{\cal Z} \rightarrow {\cal L}_G\equiv di_G +i_G d,
\label{comgenerator4}
\eeqn
where $i_G$ denotes the natural contraction of the vector field $G$
with a differential form. The BPS equation now becomes
\begin{eqnarray}
(d  - i_G)\, \Omega = \mp\, i(\delta -G\wedge )\,\Omega,
\label{bpseqinform}
\end{eqnarray}
where we use the same symbol $G$ for both the Killing
vector  and the 1-form obtained by contraction with metric.
Solving this first order system,
we should recover all 1/2 BPS and 1/4 BPS states of the underlying
Yang-Mills field theory.

\section{Angular Momentum and Eigenstates on $S^3$}

\subsection{Supermultiplet Structure of 1/4 BPS Dyons}

1/4 BPS dyons have been constructed in several difference guises. The first
was as three-pronged strings ending on D3 branes \cite{bergman},
while the field theoretical construction was as exact classical
solitons \cite{yi,hash}. Neither of these
was convenient for finding their supermultiplet structures; there are
subtleties of the respective moduli space dynamics \cite{bak,tong}.
The third method, also present in Ref.~\cite{yi}, was
a perturbative one. In this setup, one assumes very small
electric coupling and work in vacua where the 1/4 BPS dyons would be stable.
The construction proceeds by finding the lowest quantum excitation modes
around purely magnetic background. The lowest are massless moduli.
The next lowest is massive, and turns out to induce quantized
electric charges to the system when excited,
and produces many degenerate states with the same electromagnetic charges.
In the simplest case of $\balpha+\bbeta$ magnetic charge in $SU(3)$,
the total degeneracy for dyons of relative charge $q\neq 0$ is
\begin{equation}
2^6\times |2q|.
\end{equation}
Note that $q$ is quantized in half-integers. The highest spin of this
supermultiplet is $|q|+1$. The relative charge defined through the 
following expression of electric charge one may excite on the system,
\begin{equation}
{\bf q}=\left(n+q\right)\balpha +\left(n-q\right)\bbeta.
\end{equation}
where $2n$ is an integer. The consistency with Dirac quantization condition,
along with the spectrum of the original field theory,
actually demands that $n\pm q$ are integers. Thus quantization 
of $n$ is correlated with that of the relative charge $q$: a half-integral
$q$ comes with a half-integral $n$, and an integral $q$ comes with an
integral $n$.

When we reconstruct the dyons as bound states in the low energy dynamics, this 
correlation naturally emerges from the form of total moduli space 
\beq
{\cal M} = R^3 \times \frac{ R^1 \times {\cal M}_0}{Z},
\eeq
where the quotient action of $Z$ is crucial. In this note, we will not
dwell on this point. It suffices
to say that all dyonic states of relative charge $q$ can be built, 
provided that the relative part of the wavefunction is found.

Once we consider the wavefunction to be a tensor product of two parts,
one over the relative moduli space, and the other over the center-of-mass
moduli space, the degeneracy is more naturally organized as,
\begin{equation}
2^4\times\biggl((2|q|+1)+(2|q|)+(2|q|)+(2|q|-1)\biggr) .
\end{equation}
The common factor $2^4$ follows from the low energy dynamics trivially.
Among the low energy degree of freedom, there are four bosonic and
eight (real) fermionic coordinates that are associated with the center-of-mass
motion and thus free. These eight fermions acts as four pairs of
massless harmonic oscillators, whose excitations lead to $2^4=16$
degeneracy and the subsequent supermultiplet structure of $N=4$ vector
multiplet.

The rest of the degeneracy factors must arise from the relative part of the
dynamics, as we construct the 1/4 BPS dyons as bound states of monopoles.
As the above decomposition suggests, the spin content of the supermultiplet
found in Ref.~\cite{yi} is such that the bound state wavefunctions over the
relative moduli space are in
four multiplets of angular momenta; one with $l=|q|$, two with $l=|q|-1/2$,
the last with $l=|q|-1$. Thus, finding these BPS bound states
explicitly presupposes detailed understanding of
angular momentum in the low energy
dynamics.

\subsection{Isometries and Symmetries}

Let us recall the geometry of the Taub-NUT manifold. The metric is
\begin{equation}
\left(1+\frac{1}{r}\right)\,[dr^2+r^2\sigma_1^2+r^2\sigma_2^2]+\frac{1}{1+1/r}
\,\sigma_3^2,
\end{equation}
where the $\sigma_a$'s are 1-form frame on $S^3$ as in Section 2. The Taub-NUT
manifold has four Killing vectors, three of which generate $SU(2)$
rotation of $S^3$. These $SU(2)$ Killing vectors, which we denote by $L_a$,
are \cite{gibbons},
\begin{eqnarray}
&& L_1 =-\sin\phi \,\partial_\theta -
\cot\theta \cos\phi \,\partial_\phi +
{\cos\phi\over \sin\theta}\,\partial_\chi, \\
&&  L_2 = +\cos\phi \,\partial_\theta -
\cot\theta \sin\phi \,\partial_\phi +
{\sin\phi\over \sin\theta}\,\partial_\chi ,\\
&& L_3= \,\partial_\phi. \label{bosonicgenerator}
\end{eqnarray}
The $\sigma_a$'s are easily seen to be invariant under the action of
these vector fields:
\begin{equation}
{\cal L}_{L_a}(\sigma_b)=0.
\end{equation}
The operator $J_a=-i{\cal L}_{L_a}$ satisfies the usual $SU(2)$ algebra.

The fourth Killing vector corresponding to internal $U(1)$ gauge
rotations of monopoles, to be denoted as $K$, is
\begin{equation}
K=\partial_\chi.
\end{equation}
This is precisely the triholomorphic Killing vector field that enters the
low energy dynamics. This vector field do rotate $\sigma_1$ and $\sigma_2$
among themselves,
\begin{equation}
-i{\cal L}_K (\sigma_1\pm i\sigma_2)= \pm \,(\sigma_1\pm i\sigma_2),
\end{equation}
but the Taub-NUT metric itself is invariant under such rotations. Since the
range of $\chi$ is $4\pi$, the eigenvalue of $-i{\cal L}_K$ is quantized
at half-integers. Its origin as a gauge rotation generator also tells us that
its eigenvalue should be identified with the relative charge $q$
\cite{lwy}.

These isometries of the Taub-NUT manifold naturally generate symmetries of
the low energy dynamics. In the absence of potential terms due to $G=aK$,
this is especially clear since the Lagrangian is completely determined
by the metric alone. Furthermore, the geometrical interpretation of the
wavefunction $|\Omega\rangle$ as differential form $\Omega$ suggests
that these symmetries act on $\Omega$ geometrically;  In fact, it
is easy to see that the Noether charge associated with each of
these isometries do acts on wavefunctions/differential forms as Lie
derivative. Thus, for the rotational $SU(2)$, the above operator $J_a$
may be regarded as the symmetry generators. And so is $-i{\cal L}_K$
for the relative $U(1)$ gauge symmetry. 

Symmetries generated by these vectors remain symmetries of the low energy
dynamics when we turn on the potential term determined by $G=aK$. This is
guaranteed because $L_a$'s and $K$ preserves $G$:
\begin{equation}
J_a(G)=0={\cal L}_K (G).
\end{equation}
However, the $J_a$'s are not quite the physical angular momentum, it turns out.
We will come back to this crucial point shortly.

\subsection{Angular Momentum on $S^3$ and Spin}

As we observed from the known degeneracy of 1/4 BPS dyons, the bound state with
relative charge $q$ must have angular momentum that scales with
$q$ linearly. How do we realized such a multiplet? Note that the Taub-NUT
space has the topology of $R^4=R^+\times S^3$. On the $S^3$, the $SU(2)$,
which
itself is topologically $S^3$, acts freely as translations, so functions on
$S^3$ naturally fall under various $SU(2)$ representations. The angular
momentum eigenfunctions on $S^3$ are well-known to those familiar
with classic angular momentum theory, and are often denoted by $D^j_{mk}$
\cite{edmond}.
Abstractly it is defined as a finite rotation operator sandwiched between
a pair of eigenstates of total angular momentum $j$,
\begin{equation}
D^j_{mk}(\theta,\phi,\chi)\equiv \sqrt{2j+1}\;
\langle j;k|U(\theta,\phi,\chi)|j;m\rangle,
\end{equation}
where we chose a normalization so that the norm is independent of $j,m$ or $k$.
The definition makes it clear that $m$ and $k$ are bounded below and above
by $\pm j$,
\begin{equation}
j\ge m,k\ge -j,
\end{equation}
and separated from $j$ in integer steps. For any fixed $k$ or $m$, the $2j+1$
functions indexed by $m$ or by $k$, would form a spin $j$ multiplet of
$SU(2)_L$ or $SU(2)_R$, respectively, where the isometry group of undeformed
$S^3$ is $SO(4)=SU(2)_L\times SU(2)_R$.

One might worry that the angular momentum eigenstates of our low energy
dynamics may have little to do with these D-functions. After all, a
wavefunction should be regarded as a differential form whose transformation
properties are generically more complicated than functions. However,
as we pointed out earlier, the Taub-NUT manifold admits orthonormal basis
$w^\mu$, all four of which are invariant under $J_a$. This means that as
long as we construct the wavefunction/differential form in the basis
spanned by the orthonormal frame $\omega^\mu$, its transformation property
under $J_a$ originates entirely from its coefficient functions. Thus we
expect the generic form of the wavefunction could be written as,
\begin{equation}
\Omega = \sum D^j_{mk}(\theta,\phi,\chi)\;\Lambda_{j;k}(r;dr,\sigma_a),
\end{equation}
where the differential forms $\Lambda_{j;k}$ have no explicit
dependence on the three Euler angles except through the $\sigma_a$'s.
$D^j_{mk}$ with $m=-j,-j+1,\dots,j$ form a multiplet under $J_a$ with
$-i\partial_\chi$ charge $k$.

Given the definition of $D$ functions, it is not difficult to show
that the other lower indices, namely $k$, is an eigenvalue of the operator
$-i\partial_\chi$. (Because the actual geometry of the
three sphere is deformed, the $SU(2)$ under which $D^j_{mk}, k=-j,-j+1,
\dots,j$ would have formed a representation is no longer a symmetry.)
Since $j$ is bounded below by $|k|$, a large relative charge
necessarily implies a large angular momentum; The degeneracy has to scale
linearly with increasing $k\simeq q$. This is precisely the behavior we
saw from the state counting of subsection 4.1.

However, there is something missing. Given a single eigenstate
$\Omega$, we expect to generate other physical states related to it by
acting with a SUSY charge, say $Q$. The underlying field theory tells
us the new state $Q|\Omega\rangle$ must be fermionic/bosonic if
$|\Omega\rangle$ is bosonic/fermionic. In particular their physical spin
should differ by 1/2. On the other hand, $Q$ itself is invariant under the
action of $J_a=-i{\cal L}_{L_a}$, and cannot impart additional angular
momentum quantum numbers, it seems.

The resolution of this dilemma is that we should modify the angular
momentum operator by adding a ``spin'' piece. The hyperK\"ahler
structure of the Taub-NUT manifold supplies such additional conserved
quantities, fortunately \cite{vanholten}.
Let ${\cal I}^{(a)}$ be the three complex structures
as before. Define a triplet of operators $S_a$ acting on fermions by,
\begin{equation}
S_a=
\frac{i}{2}{\cal I}^{(a)}_{\mu\nu}
\varphi^\mu\varphi^{*\nu}.
\end{equation}
The pointwise action is, in geometrical terms,
\begin{eqnarray}
S_a(dz^\mu)&=&\frac{i}{2}{{\cal I}^{(a)}}^\mu_{\:\:\lambda}\;dz^\lambda, \\
S_a\left(\frac{\partial}{\partial z^\mu}\right)&=&
\frac{i}{2}{{\cal I}^{(a)}}_{\mu}^{\:\:\lambda}\;
\left(\frac{\partial}{\partial z^\lambda}\right),
\end{eqnarray}
which is nothing but the action of the three complex structures
${\cal I}^{(a)}$ up to a numerical factor. These $S_a$'s span an $SU(2)$
R-symmetry of the $N=4$ superalgebra.

The $S_a$'s themselves form a triple under the $J_a$'s, and using this fact
we may write down a new set of angular momentum generators,
\begin{equation}
M_a\equiv J_a-S_a,
\end{equation}
which also span an $SU(2)$ algebra. Note that, unlike $J_a$, $M_a$ commute
with the R-charges $S_a$.

We seem to have two possible choices of angular momentum; $J_a$ which rotate
R-charge and $M_a$ which do not. Both commute with the Hamiltonian, but
their commutators with supercharges are quite different matter. Under $J_a$,
the four complex supercharges fall into a singlet plus a triplet, i.e.
\begin{equation}
[J_a, Q]=0,\ \ \ [J_a, Q_{(b)}]=i\epsilon_{abc}Q_{(c)}\,.
\end{equation}
On the
other hand, since the four complex supercharges belong to doublets
under $S_a$, they must form doublets under $M_a$ as well. More specifically,
the following linear combinations
\begin{eqnarray}
&& Q_+=Q+iQ^{(3)},\nonumber\\
&& Q_-=iQ^{(1)}+Q^{(2)},
\label{doubleta}
\end{eqnarray}
form one doublet under $M_a$ with
$[S_3, Q_\pm]=-{1\over 2}Q_\pm$, and the second combination
\begin{eqnarray}
&& \tilde{Q}_+=iQ^{(1)}-Q^{(2)},\nonumber\\
&& \tilde{Q}_-=Q-iQ^{(3)},
\label{doubletb}
\end{eqnarray}
form another doublet under $M_a$
with $[S_3, \tilde{Q}_\pm]=+{1\over 2}\tilde{Q}_\pm$.
Since supercharges should carry physical spin 1/2, we surmise that 
$M_a$ rather than $J_a$ should be interpreted as the physical angular 
momentum. We will denote the eigenvalues of $M^2$  by $l(l+1)$.

\subsection{Anti-Self-Dual Ansatz}

Now we may proceed to write down the ansatz for dyonic BPS bound states.
The BPS equation is easily seen to be invariant under the Hodge dual operation
on the wavefunction. This property can be used to separate the self-dual part
from the anti-self-dual part of the trial wavefunction, so a BPS
wavefunction should be either self-dual or anti-self-dual.

Does the dynamics prefer one to the other?
Hamiltonian has three kinds of potential terms. In addition to the
purely bosonic potential $G^2$, there are two more terms; one is a
fermion bilinear contracted with $dG$, while the other is a fermion
quadrilinear contracted with the Riemann curvature. The salient point
is that $dG$ and the Riemann curvature are both anti-self-dual tensors
on Taub-NUT manifold. Because of this, a self-dual ansatz will
not be sensitive to some spin-spin type long range interaction,
which could be crucial for the formation of bound states. In
fact, the threshold bound state of $SU(3)$ monopoles (when $G=0$) is known
to be anti-self-dual, while no such self-dual bound state
exists \cite{lwy}. We expect that this behavior persists when $G=aK$ is turned
on, which motivates us to look for 1/4 BPS dyonic bound states with an
anti-self-dual ansatz\footnote{
Nonetheless, there is no  reason to preclude
self-dual bound states that do not saturate
the BPS bound.}

Let us start with anti-self-dual 2-forms. One interesting property of
anti-self-dual 2-forms on hyperK\"ahler 4-manifolds, is that they are
of type $(1,1)$, upon Hodge decompositions with respect to any one of
three complex structures \cite{schroer}.
Recall that, up to a numerical factor, the
``spin'' $S_a$ act on differential forms as complex structures do.
Since any form of type $(n,n)$ is annihilated by the complex
structure, we conclude that the 2-form part of an anti-self-dual
ansatz carries no spin.

On the other hand, the BPS equation connects even forms with even forms,
so an ansatz containing anti-self-dual 2-form may contain, in
addition, 0-form and 4-form. Neither carries ``spin'': 0-form
is obviously invariant under $S_a$, and a 4-form is always
proportional to the volume form which is always of type $(n,n)$. Thus, an
anti-self-dual even form is always ``spinless''; $l$ equals $j$.
The angular dependence of anti-self-dual even forms can be written
entirely in terms of $D$ functions and of the basis $\sigma_a$'s.

Of the four angular momentum multiplets, the cases of $l=q \ge 0$ and of
$l=q-1\ge 0$ belong to this category.
For $l=q$, the wavefunction should have the form,
\begin{equation}
\Omega^q_{m;q}=D^{q}_{mq}\Lambda_{q;q}+D^{q}_{m(q-1)}\Lambda_{q;q-1},
\end{equation}
for any value of $m=q,q-1,\dots,-q$, where differential forms $\Lambda$'s
can be written entirely with $r$ and $\omega^\mu$ only. This state has
relative charge $q\ge 0$ if and only if the anti-self-dual form $\Lambda$'s
satisfy
\begin{equation}
-i{\cal L}_K\Lambda_{q;q}=0,\qquad -i{\cal L}_K\Lambda_{q;q-1}
=\Lambda_{q;q-1}.
\end{equation}
Since the only charged combination one can build out of $r$ and $\omega^\mu$
is $\omega^1 \pm i \omega^2$, which has $\pm 1$ charge respectively, the
$\Lambda$'s may contribute $\pm 1$ to the total charge $q$ at most.
Furthermore, only $\omega+i\omega^2$ may enter since $D^q_{m(q+1)}$,
which should accompany $\omega- i\omega^2$, does not exist for positive $q$.

These considerations constrain possible form of $\Lambda$'s
quite severely. We find that only the following choice
is consistent with the known quantum numbers of 1/4 BPS dyons,
\begin{eqnarray}
\Lambda^{q;q}&=&f(r)+h(r)\,(\omega^0\wedge \omega^3+
\omega^1\wedge  \omega^2)+f(r)\,
(\omega^0\wedge \omega^1\wedge \omega^2\wedge\omega^3), \\
\Lambda_{q;q-1}&=&(b(r)/r)\,(\omega^0+i\omega^3)\wedge(\omega^1+i\omega^2),
\end{eqnarray}
for $l=q\ge 0$ BPS state.\footnote{We have defined the  Hodge dual operation
with respect to the volume form,
$-\omega^0\wedge\omega^1\wedge\omega^2\wedge\omega^3$.}

In case of $\,l=q-1$ with $q\ge 1$, the anti-self-dual ansatz is even
more restrictive. The only such ansatz consistent with known quantum
numbers is,
\begin{equation}
\Omega^{q-1}_{m;q}=p(r)\,D^{q-1}_{m(q-1)}\,(\omega^0+i\omega^3)
\wedge(\omega^1+i\omega^2),
\end{equation}
for $m=q-1,q-2,\dots,1-q$. No 0-form may appear since a factor of
$(\omega^1+i\omega^2)$ is necessary to
make the electric charge to be $q$, while
4-form is ruled out subsequently by the fact that the BPS
wavefunctions we are looking for are all anti-self-dual.

We will postpone discussion of the two remaining multiplets of
$l=q-1/2 \ge 0$ to the following section. The corresponding
multiplets are in 
odd forms, which can be found by acting
supercharges on $\Omega^{q}_{m;q}$ and $\Omega^{q-1}_{m;q}$.
For these two multiplets, $S_a$ contribution
to the physical angular momentum does not vanish.

\section{Dyonic BPS Bound States}

Here, we will solve for the dyonic BPS bound states explicitly. Such dyonic
bound state do not exist for all relative charge $q$. Rather, it is known
that $|q|$ must be smaller than the critical charge $|q_{cr}|$
\cite{bergman,yi,tong}, where
\begin{equation}
q_{cr}=\lim_{r\rightarrow\infty}a\langle K,K\rangle.
\end{equation}
With our current normalization, $\langle K,K\rangle$
asymptotes to $1$ at infinity, so the critical charge $q_{cr}$
is equal to the parameter $a$.
Thus we expect to find the BPS bound state, that is, a normalizable and regular
wavefunction that preserves half of low energy supersymmetry, only when
$|q|\le |a|$. Without loss of generality, we will take both $a$ and $q$ to be
nonnegative.

Classical analysis of Ref.~\cite{bergman,yi} leaves it unclear whether the
bound state should exist (at threshold) when $a=q$. Classically two 
monopoles are infinitely separated, but quantum mechanically, there 
may be a bound state with powerlike decay. As the following analysis
will show, however, no such threshold bound state exists, except for 
$a=q=0$ case.

\subsection{The $l=q-1$ Multiplet}

The case of $l=q-1$ is the simplest. Starting with the ansatz,
\begin{equation}
\Omega^{q-1}_{m;q}=p(r)\,D^{q-1}_{m,q-1}\,(\omega^0+i\omega^3)
\wedge(\omega^1+i\omega^2),
\end{equation}
the BPS equation reduces to a single ordinary differential equation for $p(r)$,
\begin{equation}
\frac{d}{dr}\,r\,p(r)=-A(r)\,r\,p(r),
\end{equation}
where the quantity $A$ is defined to be the following combination:
\begin{equation}
A\equiv a-q\left(1+\frac{1}{r}\right).
\end{equation}
A useful fact we employed is that
\begin{equation}
e^r\;(\omega^0+i\omega^3)\wedge(\omega^1+i\omega^2)
\end{equation}
is a (nonnormalizable) harmonic 2-form. The equation is easily solved to
give the wavefunction:
\begin{equation}
\Omega^{q-1}_{m;q}= D^{q-1}_{m(q-1)} {r^{q-1}e^{-(a-q)r}}\,
(\omega^0 + i\omega^3)\wedge(\omega^1 +i\omega^2).
\end{equation}
The wavefunction is exponentially small at large $r$ and
normalizable as long as $a>q$. This way, we have recovered the
fact that $a$ is the critical electric charge, beyond which no
bound state may exist.
Furthermore, the wavefunction is regular as long as $q\ge 1$, which is also
consistent with the fact that $l=q-1$ BPS bound state exists only for $q\ge 1$.

\subsection{The $l=q$ Multiplet}

The ansatz for $l=q\ge 0$ is a bit more involved;
\begin{eqnarray}
\Omega_{q;q}&=&D^{q}_{mq}\left(f(r)+h(r)\,(\omega^0\wedge
\omega^3+\omega^1\wedge  \omega^2)+f(r)\,
(\omega^0\wedge \omega^3\wedge \omega^1\wedge \omega^2) \right) \nonumber \\
&+&D^{q}_{m(q-1)}\left(
(b(r)/r)\,(\omega^0+i\omega^3)\wedge(\omega^1+i\omega^2)\right),
\end{eqnarray}
under which BPS equations reduce to
\begin{equation}
\frac{d}{dr}f= -Ah +{b\over r^2},\qquad
\frac{d}{dr}h + \frac{2h}{1+r}= -Af - {b\over r^2},\qquad
\frac{d}{dr}b +Ab=q(f-h).
\label{antiselfdual}
\end{equation}
In order to solve (\ref{antiselfdual}) for general $q>0$, we proceed as
follows. By substituting,
\begin{equation}
f=u(r) e^{-\int A},\qquad h=v(r)  e^{-\int A},\qquad
b=w(r) \left({q\over A}\right) e^{-\int A},
\end{equation}
to (\ref{antiselfdual}),
one obtains,
\begin{equation}
{d\over dr} (u-w)=0,\qquad
{d\over dr}u+{1\over (1+r)^2}{d\over dr}(1+r)^2 v =0,\qquad
{d\over dr}(w/A)=u-v \,.
\end{equation}
We solve the first equation by $u=C_1+w$ with an integration constant $C_1$.
The remaining equations can be combined into a single second order equation,
\begin{eqnarray}
\label{bps5}
{d\over dr}\left[ {d\over dr}\left({(1+r)w\over A}\right) - 2A
\left({(1+r)w\over A}\right) - 2C_1 r\right]=0,
\end{eqnarray}
which is integrated with a second integration constant $C_2$ to
\begin{eqnarray}
\label{bps6}
{d\over dr}\left({(1+r)w\over A}\right) - 2A
\left({(1+r)w\over A}\right) =+ 2C_1 r - 2 C_2 .
\end{eqnarray}
Integrating this equation gives us three-parameter family of solutions.
But, fortunately, the simplest possible solution, 
\begin{eqnarray}
\label{solution}
{(1+r)w\over A} =- r,
\end{eqnarray}
turns out to be the only regular and normalizable solution. Using it to
generate other radial functions, we find the $l=q$ bound states:
\begin{eqnarray}
\Omega^q_{m;q}&=&D^q_{mq} {r^qe^{ - (a-q)r}\over 1+r}\,\left[
a +  (a+ {1\over 1+r})( \omega^0\wedge\omega^3 +
\omega^1\wedge\omega^2) +
a\,\omega^0\wedge\omega^1\wedge\omega^2\wedge\omega^3)
\right]\nonumber\\
 &-& D^q_{m(q\!-\!1)}{r^qe^{ -(a-q)r}\over 1+r}\,\sqrt{q/2}\,
(\omega^0 + i\omega^3)\wedge (\omega^1 +i\omega^2).
\end{eqnarray}
Again, we find that the wavefunction is normalizable as long as $a>q$.
The solution is regular at origin for all nonnegative $q$. We have
recovered the $l=q$ multiplet of 1/4 BPS dyon of charge $q>0$.

The case of $q=0$ is a bit special, where the BPS state is a purely magnetic
bound state of the two monopoles. For $q=0$, the second $D$ function
does not exist, and we must solve a modified BPS equation. Nevertheless,
the actual wavefunction is also obtained by taking $q=0$ limit of the above
result:
\begin{equation}
\Omega^0_{0;0}={e^{ - a r}\over 1+r}\,\left[
a +  (a+ {1\over 1+r})( \omega^0\wedge\omega^3 +
\omega^1\wedge\omega^2) +
a\,\omega^0\wedge\omega^1\wedge\omega^2\wedge\omega^3)
\right].
\end{equation}
This BPS state actually preserves all supercharges of low energy dynamics.
In fact, this is the lowest lying state of this low energy effective
theory. In the limit of aligned vacua ($a=0$), this state also
reverts to the threshold bound state of two monopoles found in Ref.~\cite{lwy},
as it should.

\subsection{The $l=q-1/2$ Multiplets}

The remaining two multiplets of $l=q-1/2$ can be found most easily
by acting $Q$ on $\Omega^q_{m;q}$ and $\Omega^{q-1}_{m;q}$ found above.
{}From  $\Omega^{q-1}_{m;q}$, we find $2q-1$ states
\begin{eqnarray}
\label{oddsolb}
{r^qe^{ - (a-q)r}\over\sqrt{r+r^2}}\,(\omega^1 +
i\omega^2)\wedge(1+\omega^0\wedge\omega^3)D^{q-1}_{m(q-1)},
\end{eqnarray}
while $\Omega^{q}_{m;q}$ produces $2q+1$ states
\begin{eqnarray}
\label{oddsola}
{r^qe^{ - (a-q)r}\over \sqrt{r+r^2}}\,\Bigl[ (\omega^0 \!+\!
i\omega^3)\wedge(1\!+\!\omega^1\wedge\omega^2) \sqrt{2q}D^q_{mq}+ i
(\omega^1 \!+\!
i\omega^2)\wedge(1\!+\!\omega^0\wedge\omega^3)D^q_{m(q\!-\!1)} \Bigr].
\end{eqnarray}
These account for all $4q$ states in the two $l=q-1/2$ multiplets.
(Alternatively we could have used $Q^{(a)}$ instead of $Q$. Since
$\Omega^{q-1}_{m;q}$ and $\Omega^{q}_{m;q}$ are invariant
under complex structures, and since $Q^{(a)}$ are essentially $Q$ rotated
by the complex structures, the resulting $4q$ states will be simply
the above $4q$ states with complex structure ${\cal I}_a$ acting on
them.)

However, because $Q$ is a singlet under $J_a$, the above states form
multiplets under $J_a$ instead of the physical angular momentum $M_a$.
To reconstruct $M_a$ multiplets, we recall the relationship
\begin{eqnarray}
J_a=M_a+S_a
\end{eqnarray}
Since the two $SU(2)$ generators, $M_a$ and $S_a$, commute with each other,
$J_a$ multiplets are constructed from $M$ multiplets
and $S$ multiplets by the rule of
angular momentum addition. On the other hand, we actually need to
reconstruct $M_a$ eigenstates from $J_a$ and $S_a$ eigenstates, for
which we need to reverse the procedure. Without delving into details
of the computation, we present the two $l=q-{1\over 2}$ multiplets.
Because $S_3$ commutes with $M_a$, we can label the two multiplets
by its eigenvalue $s_3$. The first has $s_3=1/2$;
\begin{eqnarray}
&&{(\Omega_{(+)})}^{(q\!-\!{1\over 2})}_{m;q}= {r^q e^{ - (a-q)r}\over
\sqrt{r+r^2}}\Bigl[ (\omega^0 \!+\! i\omega^3)\wedge
(1\!+\!\omega^1\wedge\omega^2)\sqrt{q\!+\!m\!+\!{1\over 2}\over 2q+1}\,
D^q_{(m+{1\over 2})q}\nonumber\\ &&+i (\omega^1
\!+\! i\omega^2)\wedge(1\!+\!\omega^0\wedge\omega^3) \Bigl(
\sqrt{q\!+\!m\!+\!{1\over 2}\over
2q(2q+1)}D^q_{(m\!+\!{1\over 2})(q\!-\!1)}
\!+\!\sqrt{q\!-\!m\!-\!{1\over 2}\over 2q}D^{q-1}_{(m\!+\!{1\over 2})
(q\!-\!1)}\Bigr) \Bigr] \,. \label{spinup}
\end{eqnarray}
The other multiplet has $s_3=-1/2$;
\begin{eqnarray}
&&{(\Omega_{(-)})}^{(q\!-\!{1\over 2})}_{m;q}= {r^qe^{ - (a-q)r}\over
\sqrt{r+r^2}}\Bigl[ (\omega^0 \!+\! i\omega^3)\wedge
(1\!+\!\omega^1\wedge\omega^2) \sqrt{q\!-\!m\!+\!{1\over 2}\over 2q+1}
D^q_{(m\!-\!{1\over 2})q}\nonumber\\ &&+i
(\omega^1 \!+\! i\omega^2)\wedge(1\!+\!\omega^0\wedge\omega^3) \Bigl(
\sqrt{q\!-\!m\!+\!{1\over 2}\over 2q(2q+1)}D^q_{ (m\!-\!{1\over
2})(q\!-\!1)} \!-\!\sqrt{q\!+\!m\!-\!{1\over 2}\over 2q}D^{q-1}_{
(m\!-\!{1\over 2})(q\!-\!1)}\Bigr) \Bigr] \,. \label{spindown}
\end{eqnarray}
In both expressions, the index $m$ takes values $q-1/2,q-3/2,\dots,-q+1/2$.

In a direct construction of these two $l=q-{1\over 2}$ multiplets
from  the $l=q$ or $l=q-1$ multiplets, the role of the doublet supercharges 
in (\ref{doubleta}) and (\ref{doubletb})  can be easily identified.
As a simple application of the  angular momentum addition rule, the 
operations of the doublets on the even-form multiplets will produce
$(q+{1\over 2})\oplus (q-{1\over 2})\,\,\Bigl[={1\over 2}\otimes q\Bigr]$ or
$(q-{1\over 2})\oplus (q-{3\over 2})
\,\,\Bigl[={1\over 2}\otimes (q\!-\!1)\Bigr]$,
but one may check that both $l=q+{1\over 2}$ and
$l=q-{3\over 2}$ multiplets vanish
identically.
Since the doublet $Q_\pm$ carries the spin eigenvalue
$s_3=-{1\over 2}$, the operation of
this doublet
on $l=q$ or $l=q-1$ multiplets, produces
the $l=q-{1\over 2}$
multiplet with  $s_3=-{1\over 2}$.
Similarly, the application of
$\tilde{Q}_{\pm}$ results in the  $l=q-{1\over 2}$
multiplet with  $s_3=+{1\over 2}$. 

\subsection{Characteristics of the BPS states}
In the construction of the BPS bound states,
we have limited ourselves to the case of nonnegative
electric charge.
For negatively charged bound states, we
note the fact that the complex conjugation of  a solution to
the BPS
equations in (\ref{bpseqinform}), gives another
solution.
Both the eigenvalues of  charge and  $M_3$
reverse their signs under the complex conjugation.
Thus the negatively charged solution,
 $\Omega^l_{\!-m;\!-q}\ (q\ge 0)$
is simply given by the complex conjugation of
$\Omega^l_{m;q}$.

We now turn
to the case that the Higgs misalignment
parameter $a$ is negative.
When $a$ is replaced by $-a$,
only the fermionic term that couples to the
Killing potential changes its sign
in the supersymmetric Lagrangian (\ref{action}). By the parity
operation $\psi\rightarrow  i\gamma^0 \psi$ or equivalently
$\varphi\rightarrow i\varphi$, one can
bring this Lagrangian  to the original form
with $a>0$. The corresponding  
transformation amounts to replacement of
$dx^\mu$ by $idx^\mu$. The solutions for $-a$
are obtained if one replaces
all $\omega^\mu$ of the above BPS solutions
by $i\omega^\mu$. These exhaust all
the possibilities.

For the remainder of the section, we would like to comment
briefly on some other aspects of the BPS states. With an excited charge $q$, 
the effective potential at large relative separation tends to $(a^2+q^2)/2$. 
Since the energy eigenvalue of the BPS multiplets is $|aq|$, one finds that
binding energy of the dyons is
\begin{eqnarray}
E_{\rm binding}={(|a|-|q|)^2\over 2} ,
\end{eqnarray}
which tends to zero as the charge approach its critical value, as one
expects from its classical counterpart.
Another characteristic is the separation of the two monopole cores. 
In the classical limit, the separation between the two
cores are given by $r_{\rm eq}\equiv |q|/(|a|-|q|)$. 
We expect the vacuum expectation value,
\begin{equation}
\langle r \rangle \equiv \frac{\langle\Omega|r
|\Omega\rangle}{\langle\Omega|\Omega\rangle}
\end{equation}
to approach $r_{\rm eq}$ in the classical limit. For a given supermultiplet 
with charge $q$, the expectation values are found to be dependent
upon the  angular momentum quantum number $l$. For instance, 
we find the expectation value
\begin{eqnarray}
\langle r\rangle =
{|q|\over (|a|-|q|)}\left(1+{1\over 2|a|}\right)=r_{\rm eq}\,
\left(1+{1\over 2|a|}\right)\,,
\end{eqnarray}
for the $l=|q|-1$ multiplet, and
\begin{eqnarray}
\langle r\rangle =
{|q|\over (|a|-|q|)}\left(1+{1\over 2|q|}\right)=r_{\rm eq}\,
\left(1+{1\over 2|q|}\right)\,,
\end{eqnarray}
for the $l=|q|-1/2$ multiplets. To restore the Planck constant $\hbar$,
we simply observe that classical charge $q$ has the same dimension as 
$\sqrt\hbar$. Since the difference between $ \langle r\rangle $ and 
$r_{\rm eq}$ scales inversely with $q$ or $a=q_{cr}$, it has to scale
linearly with $\sqrt{\hbar}$:
\begin{equation}
\frac{\langle r\rangle - r_{\rm eq}}{r_{\rm eq}}\sim O({\sqrt\hbar}) \;.
\end{equation}
Thus, $\langle r\rangle $ indeed
approaches $r_{\rm eq}=|q|/(|a|-|q|)$ in the classical limit.

\section{Conclusion}

In the low energy dynamics of 1/2 BPS monopoles, the misaligned Higgs vacua
induces an attractive potential between monopoles of distinct types. This
potential is crucial in formation of new dyonic bound states of monopoles,
some of which preserve 1/4 of supersymmetries in field theoretical sense,
or equivalently, 1/2 of supersymmetries of the low energy dynamics of
monopoles. Starting from the full N=4 supersymmetric low energy effective
Lagrangian, we expressed the BPS equation of the system in the language
of the differential form, which was then further reduced to a set of coupled
first-order ordinary differential equations. These equations were
solved analytically, giving 1/4 BPS dyons
as quantum bound states of two distinct SU(3) monopoles. Along the course
of the construction, we have given  a full account of the
supermultiplet structures of the quantum 1/4 BPS dyons.

In this note, we have focused on the BPS saturated
states in pursuit of the 1/4 BPS dyons. However,
it is expected that there exist spectra of other dyonic bound
states  that do not saturate the BPS bound.
Problem of finding these non-BPS bound states should be
quite involved. Nevertheless, there are some additional information
that might be of help. Supersymmetric sigma models on the Taub-NUT 
geometry are known to allow additional conserved quantities
of the Runge-Lenz type \cite{gibbons,vanholten}. It seems quite plausible
that this new symmetry generalize to the present low energy dynamics
with potential. We have checked that the purely bosonic part indeed
admits such conserved quantities. Such additional symmetries
might be useful in finding the excited non-BPS bound states.

Another aspect of the dynamics we did not discuss here is the scattering 
of dyons. The supersymmetric quantum mechanics we used can be thought 
of low energy dynamics for 1/4 BPS dyons in two ways. First, it produces
these dyons as bound states. Second, it provides a framework where
interaction among these dyons can be studied in a quantum mechanical setting.
The dynamics is admittedly more involved than the usual moduli space
dynamics, given the presence of the potential. It requires further study.

\vspace{3mm}

\centerline{\bf Acknowledgments}

D.B. is supported in part by Ministry of
Education Grant 98-015-D00061.  K.L.  are supported in part
by the SRC program of the SNU-CTP and the Basic Science and Research
Program under BRSI-98-2418.  D.B. and K.L. are also supported in part
by KOSEF 1998 Interdisciplinary Research Grant 98-07-02-07-01-5.
K.L. acknowledges Aspen Center for Physics where this work was completed.

\end{document}